\documentclass[prb,twocolumn,showpacs,preprintnumbers,amsmath,amssymb]{revtex4}
\usepackage{graphicx}
\usepackage{dcolumn}
\usepackage{bm}
\usepackage[tight]{subfigure}
\usepackage{enumerate}
\usepackage{amsmath}
\usepackage{verbatim}
\usepackage{color}
\usepackage{lipsum}

\begin{document}
\title{Nonlinear thermoelectric efficiency of superlattice-structured nanowires}
\author{Hossein Karbaschi$^{1,2}$}
\author{John Lov\'en$^{1}$}
\author{Klara Courteaut$^{3}$}
\author{Andreas Wacker$^{3}$}
\author{Martin Leijnse$^{1}$}

\affiliation{$^{1}$Division of Solid State Physics and NanoLund, Lund University, Box 118,S-221 00 Lund, Sweden}
\affiliation{$^{2}$Department of Physics, Faculty of Sciences, University of Isfahan, Isfahan 81746-73441, Iran} 
\affiliation{$^{3}$Division of Mathematical Physics and NanoLund, Lund University, Box 118,S-221 00 Lund, Sweden} 

\begin{abstract}
We theoretically investigate nonlinear ballistic thermoelectric transport in a superlattice-structured nanowire.
By a special choice of nonuniform widths of the superlattice barriers---analogous to anti-reflection coating in optical systems---it 
is possible to achieve a transmission which comes close to a square profile as a function of energy. 
We calculate the low-temperature output power and power-conversion efficiency of a thermoelectric generator based on such a structure 
and show that the efficiency remains high  also when operating 
at a significant power. To provide guidelines for experiments, we study how the results depend on the nanowire radius, the number 
of barriers, and on random imperfections in barrier width and separation. Our results indicate that high efficiencies can indeed 
be achieved with todays capabilities in epitaxial nanowire growth. 

\end{abstract}
\maketitle

\section{Introduction}
In a thermoelectric material, an applied heat gradient gives rise to an electric current or voltage and, conversely,
an applied electric current gives rise to a temperature difference. Therefore, thermoelectric devices can be used as 
power generators, converting heat into electric power, or as refrigerators, using electric power to accomplish cooling~\cite{RoweTEbook, Goldsmid}.
However, the efficiency of todays thermoelectric devices is much lower than that of the alternatives (mostly using the compression and expansion of gases) 
and their use is therefore 
limited to a number of niche applications where advantages such as small size, no moving mechanical parts, 
reliability, and capability to generate power also at small heat gradients give them the upper hand. 
Many different ways to enhance the efficiency of thermoelectric devices have been proposed, the most relevant for the 
work we present here being nanoscaling, i.e., the idea to reduce the dimensions of devices or introduce structuring on very 
small length scales, which was first investigated by Hicks and Dresselhaus~\cite{Hicks93a, Hicks93b} and has now seen some experimental 
success, see e.g., Refs.~\citenum{Venkatasubramanian01, Harman2002, Wu13}. 

A question of more interest for fundamental physics is how good you can possibly make a thermoelectric device. Focusing on a 
power generator for definiteness, the laws of thermodynamics sets a fundamental limit on the efficiency of any device 
extracting work by using the temperature difference between a hot bath (temperature $T_h$) and a cold bath (temperature $T_c$), 
which can never be higher than the Carnot efficiency, $\eta_C = 1-T_h/T_c$.
It was shown in Refs.~\citenum{Mahan96, Humphrey05} that a thermoelectric device can indeed, in theory, operate at 
the Carnot efficiency. This is only possible for a 
material where electrons can only be transported at one particular energy, meaning that the 
transmission function (for ballistic transport) or transport distribution function (for diffusive transport) has to be proportional 
to the delta function. It has later been noted that quantum dots~\cite{Esposito09} or molecules~\cite{Murphy08, Leijnse10} weakly coupled 
to leads would possess precisely such electronic transport properties due to the discrete orbitals.

However, it has later been noted that discrete orbital states are not ideal when trying to 
operate a thermoelectric device at large output power, because this requires a large tunnel coupling to the leads which 
broadens the orbitals and spoils the delta-function like transmission~\cite{Nakpathomkun10, Karlstrom11}. 
Recently, Refs.~\citenum{Whitney14, Whitney15} addressed, 
and solved, the problem of finding the transmission function which maximizes the efficiency at a given desired output power, showing that it 
should have a square shape (i.e., letting all electrons through within a finite energy window and blocking all transport 
outside this window). 


In this work, we theoretically investigate a superlattice-structured semiconductor nanowire (NW) in the regime of ballistic 
transport, and show that this is a possible realization of a system with a nearly perfectly square-shaped 
transmission function.
The electron contribution to both the charge current and heat current is calculated taking the full voltage dependence 
and nonequilibrium condition into account, but neglecting the phonon contribution to the heat current (which would simply be an additive 
loss-mechanism), inelastic scattering, and electron-electron interactions.
In close analogy with anti-reflection coating used in optical systems, a special choice of non-uniform barrier widths makes it 
possible to achieve energy windows with almost perfect 
transmission or reflections of electrons~\cite{Pacher01}, even with a rather small number of barriers.
The electronic power-conversion efficiency can then come rather close to the 
Carnot efficiency, and the efficiency remains large also when operating at large output powers. To investigate the demands our 
proposal sets on epitaxial NW growth, we investigate how the results depend on the number of barriers, showing that a rather 
small number is sufficient, as well as the sensitivity to random imperfections in the barrier width and separation. We 
here use parameters valid for InAs NWs, where barriers can be formed by controllable growth of InP segments~\cite{Bjork04} or of 
segments of wurtzite structure in an otherwise zinc blende NW~\cite{Zanolli07, Belabbes12, Namazi15, Nilsson16}.

Previous studies have investigated thermoelectric effects in superlattice-structured NWs in the linear response 
regime, see e.g., Refs.~\citenum{Lin03, Chen2011, Shi2012, Wang14}, but, to the best of our knowledge, there have been no systematic attempts to 
design a square transmission function, or to investigate the nonlinear regime. 

The paper is organized as follows. 
Section~\ref{sec:model} introduces the NW model and explains the simple theory for ballistic thermoelectric transport 
used throughout this work. 
Section~\ref{sec:results} contain the main results. In Sect.~\ref{sec:res_transmission} we discuss the transmission function 
and show how to design the barriers to achieve a good square shape even with a rather small number of barriers.  
Section~\ref{sec:res_generator} shows the resulting nonlinear power and efficiency of the device operated as a thermoelectric 
power generator.
For simplicity, Sects.~\ref{sec:res_transmission} and~\ref{sec:res_generator} focus on the ideal case of a single transport channel 
and perfectly uniform superlattice parameters; Sec.~\ref{sec:res_disorder} then collects the discussions of deviations from the 
ideal situation. Finally, Sect.~\ref{sec:conclusions} concludes and summarizes our findings.  

\section{Nanowire model and thermoelectric transport \label{sec:model}}
\begin{figure}[]
  	\includegraphics[height=0.75\linewidth]{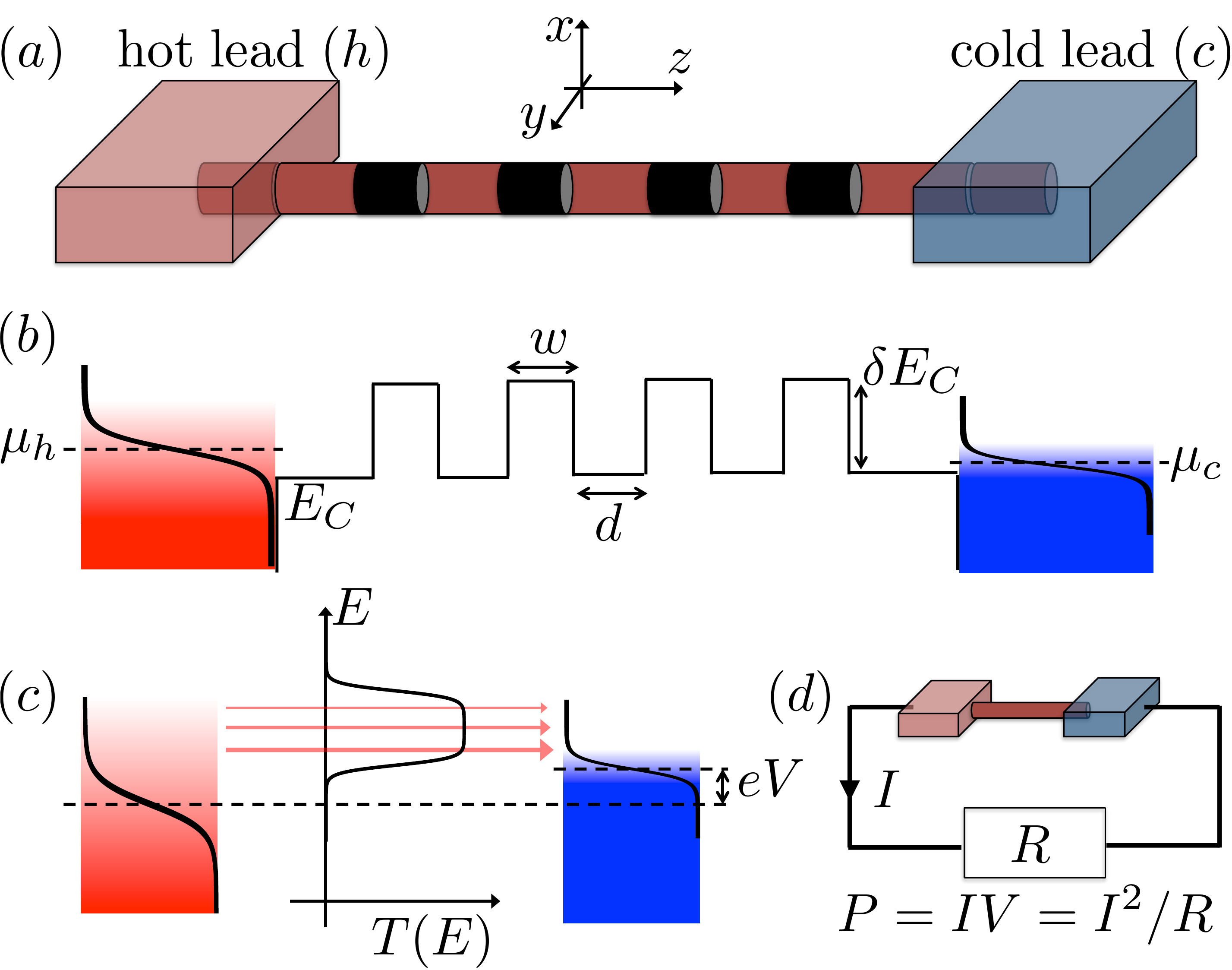}	
	\caption{\label{fig:setup}(Color online) 
	(a) Sketch of setup with a superlattice-structured NW placed in the $z$-direction and connected to a hot and a cold lead.
	(b) Energy diagram corresponding to the setup in (a). 
	(c) Energy diagram where the NW is represented by its transmission function $T(E, V=\mathrm{const})$. 
	(d) Sketch showing how the thermoelectrically generated power can be used in an external circuit (here simply 
	represented by a resistor), which generates the voltage $eV = \mu_c - \mu_h$ across the thermoelectric element 
	in (b) and (c).}
\end{figure}

Figure~\ref{fig:setup}(a) shows a sketch of the considered device with a superlattice-structured NW attached to a 
hot and a cold lead, where the barriers are indicated by black NW segments.
Figure~\ref{fig:setup}(b) is a mixed real space and energy space sketch of the structure in (a), where the hot and cold leads 
are represented by Fermi seas of electrons with temperatures $T_h$ and $T_c$, and electrochemical potentials $\mu_h$ and $\mu_c$, 
respectively (the thermally broadened electron distributions are indicated both by the color grandient and by a curved 
line representing the value of the Fermi function).  
The conductance band edge of the NW, $E_C$, is also shown, which is increased in energy by an amount 
$\delta E_C$ at the barriers, which have thickness $w$ and separation $d$.  

Our treatment is based on a coherent evolution along the superlattice
neglecting dephasing.  This is realistic at cryogenic temperature for
superlattices up to 10 barriers as demonstrated in Ref.~\citenum{Rauch98}.
In this ballistic transport regime, the electric current is given by 
\begin{equation}\label{eq:current}
	I = \frac{2 e}{h} \int dE \; T(E, V) \left[ f\left(\frac{E-\mu_h}{k_B T_h}\right)-f\left(\frac{E-\mu_c}{k_B T_c}\right) \right],
\end{equation}
where we for simplicity have focused on the strictly one-dimensional (1D) case of a single transport channel 
[see Eqs.~(\ref{eq:current_multi}) and~(\ref{eq:current_3D}) for the expressions for multiple 1D channels and 3D devices, 
respectively]. In Eq.~(\ref{eq:current}), $-e$ is the electron charge, $h$ is Planck's constant, $V = (\mu_c - \mu_h)/e$ is the bias 
voltage, $k_B$ is Boltzmann's constant, $f(x) = 1/(\mathrm{exp}(x)+1)$ is the Fermi function, and
$T(E, V)$ is the transmission function (which depends explicitly on bias voltage).

The thermoelectric effect in a ballistic conductor is most easily understood by considering first a temperature difference between the 
two leads, $T_h > T_c$, while keeping the electrochemical potentials equal, $\mu_h = \mu_c = \mu$. If the transmission function close 
to $\mu$ is energy independent the imbalance in state occupations in the leads will give rise to a current 
of high-energy electrons with $E > \mu$ flowing from the hot to the cold lead, and an equally large current of low-energy electrons 
with $E < \mu$ flowing in the opposite direction. Therefore, there will be a net heat current $Q > 0$ flowing from hot to cold, but no net 
electric current, $I = 0$. An electric current results if the transmission function is asymmetric around $E=\mu$, for example letting only the 
high-energy electrons through as in Fig.~\ref{fig:setup}(c). Electric power, $P = IV$, now results if 
the electric current is driven "up stream", against the voltage. In a thermoelectric power generator, the voltage is generated by the 
external electric circuit where the current does electric work (or charges a battery), represented in Fig.~\ref{fig:setup}(d) by a 
resistor. Here we will for simplicity instead consider $V$ to be externally controllable. The efficiency of the thermoelectric generator is now given by 
the generated electric power divided by the heat which is lost from the hot lead 
\begin{equation}\label{eq:eta}
	\eta = \frac{P}{Q_h}.
\end{equation}
We will neglect all other contributions to $Q_h$ except the heat carried by the electrons. It is important to realize that even though
we neglect inelastic scattering of the electrons traversing the device between the hot and cold leads, there is no conservation of 
heat current, $Q_h \neq -Q_c$. Instead, the first law of thermodynamics gives $P = Q_h + Q_c$ (with $Q_c < 0$) and Eq.~(\ref{eq:eta}) 
can equivalently be written as $\eta = 1 - |Q_c|/Q_h$.

The 1D expression for the electron contribution to the heat current flowing out of lead $h$ is  
\begin{eqnarray}\label{eq:heatcurrent}
	Q_h &=& \frac{2}{h} \int dE \; T(E, V) (E - \mu_h)  \nonumber \\ 
	    &\times& \left[ f\left(\frac{E-\mu_h}{k_B T_h}\right) - f\left(\frac{E-\mu_c}{k_B T_c}\right) \right],
\end{eqnarray}
i.e., the same as Eq.~(\ref{eq:current}) but replacing the electric charge of the electrons with the heat $E - \mu_h$ they carry
[see Eqs.~(\ref{eq:heatcurrent_multi}) and~(\ref{eq:heatcurrent_3D}) for the corresponding expressions for multiple 1D channels and 3D devices, 
respectively].
We can now understand the optimal shape of the transmission function. To maximize $P$, we should allow electrons to travel only 
in one direction, for example from hot to cold, which requires a strong energy asymmetry in the transmission function. 
To minimize $Q_h$, we should, on the other hand, only let electrons through close to $\mu_h$. Therefore, the maximal $\eta$ is 
obtained when $T(E, V)$ is maximally peaked in $E$, i.e., proportional to $\delta(E)$~\cite{Mahan96, Humphrey05}.
However, a finite number of transport channels gives rise to a peak of 
finite height (rather than a true delta function) and $I$, and therefore $P$, approaches zero as this peaks becomes increasingly narrow. 
It was shown in Refs.~\citenum{Whitney14, Whitney15} that the optimal $\eta$ at a given $P$ is instead 
obtained when $T(E, V)$ at a fixed $V$ has a square shape, allowing all electrons to be transmitted within an energy range determined by the desired 
efficiency and blocking all electrons outside this energy range.   

A sharp onset in the transmission function can be achieved by using doping or electrostatic gating to position the lowest 
1D subband of a NW close to $\mu_{h,c}$ (this idea was originally suggested by Hicks and Dresselhaus~\cite{Hicks93b}, although
they considered diffusive rather than ballistic transport). Experimental realization of this proposal has been difficult, e.g., 
because disorder introduces scattering which ruins the sharp onset of the transmission as the lowest subband falls below $\mu_{h,c}$ 
(interestingly though, the resulting rapid fluctuations in the transmission close to pinch-off was recently shown to in some cases give rise 
to an increased $P$ compared with the bulk material~\cite{Wu13}).   
In the superlattice structure we consider here, we can allow $\mu_{h,c}$ to lie well within the lowest subband, 
because the quantum mechanical reflection from the barriers
block transport except for within narrow energy bands. 

We use a continuum model and want to calculate the current due to plane waves travelling in the $z$-direction, i.e., we 
consider wavefunctions of the form $\Psi(\mathbf{r}) = \phi_{n}(x,y) \psi(z)$, where $\phi_{n}(x,y)$ is the radial part and 
\begin{equation}\label{eq:psi}
\psi(z) = A e^{i k_z z} + B e^{-i k_z z},
\end{equation}
where $k_z = \sqrt{2m_z^*(E - E_C - E_{n})}/\hbar$, with $m_z^*$ the effective mass in the $z$-direction 
and $E_{n}$ the energy of subband $n$ (measured relative to $E_C$).
By matching the wavefunctions at points where the potential changes, we find the T-matrix $\mathcal{T}^{12}$ which 
relates the coefficients of the rightgoing and leftgoing waves to the left of the potential change ($A_1$ and $B_1$) 
to those to the right of the potential change ($A_2$ and $B_2$)~\cite{Daviesbook}
\begin{equation}
\left( 
\begin{array}{c}
A_1 \\ B_1
\end{array}
\right) = 
\mathcal{T}^{12}
\left( 
\begin{array}{c}
A_2 \\ B_2
\end{array}
\right). 
\end{equation}
The T-matrix for a more complicated structure involving $N$ points where the potential changes is simply found by multiplying the 
individual T-matrices, $\mathcal{T}^{1N} = \mathcal{T}^{12} \mathcal{T}^{23} \hdots \mathcal{T}^{(N-1)N}$ and the total transmission
amplitude is found from the matrix elements, 
$t = (\mathcal{T}^{1N}_{11} \mathcal{T}^{1N}_{22} - \mathcal{T}^{1N}_{12}\mathcal{T}^{1N}_{21})/ \mathcal{T}^{1N}_{22}$. Finally, 
the transmission function is given by $T(E, V) = (k_{z,N}/k_{z, 1}) |t|^2$ (assuming equal effective mass in regions $1$ and $N$).  

The voltage dependence in $T(E,V)$ originates from the variation in the potential profile along the wire when $V > 0$. 
For simplicity, we assume that the potential changes only in the barriers and remains constant in between, such that the total 
potential drop over the entire structure is equal to $eV$ (when the barriers have varying width we assume that the voltage drop 
over each barrier is proportional to its width). In addition, we neglect the small slope at the top of the barriers and assume them to 
remain square shaped. A more rigorous approach would be to self-consistently solve for the potential profile and charge density along the 
wire, but we do not expect significant deviations from our simpler method because we focus on small voltages and chemical 
potentials below the transmission band, and therefore small charge densities. 

\section{Results \label{sec:results}}
To make the results as easy as possible to understand, we focus in Sects.~\ref{sec:res_transmission} 
and~\ref{sec:res_generator} on a fully 1D NW, 
meaning that all subbands except the lowest lies far above $\mu_h$ and $\mu_c$, and we assume a perfect structure (equal height, 
width, and spacings of all barriers). Deviations from these conditions will be investigated in Sect.~\ref{sec:res_disorder}.

Unless otherwise is stated, we use $T_c = 4~\mathrm{K}$, $T_h = 10~\mathrm{K}$, and $m^* / m_0 = 0.022$, 
where $m_0$ is the free electron mass and $m^*$ is the effective mass appropriate for the conduction band of InAs. 
We also assume $\delta E_C = 100~\mathrm{meV}$, which is close to the value extracted from recent 
experiments~\cite{Nilsson16} for the barrier height associated with wurtzite segments in a zinc blende InAs NW. 
Furthermore, we use $w=15~\mathrm{nm}$, $d=10~\mathrm{nm}$, and, unless otherwise is stated, 7 barriers. 
To obtain a square-shaped transmission with a small number of barriers we follow
Ref.~\citenum{Pacher01} and let the two outermost barriers be half the width of the others, analogous to anti-reflection 
coating in optics. 

\subsection{Optimizing the transmission function \label{sec:res_transmission}}
\begin{figure}[]
  	\includegraphics[height=0.8\linewidth]{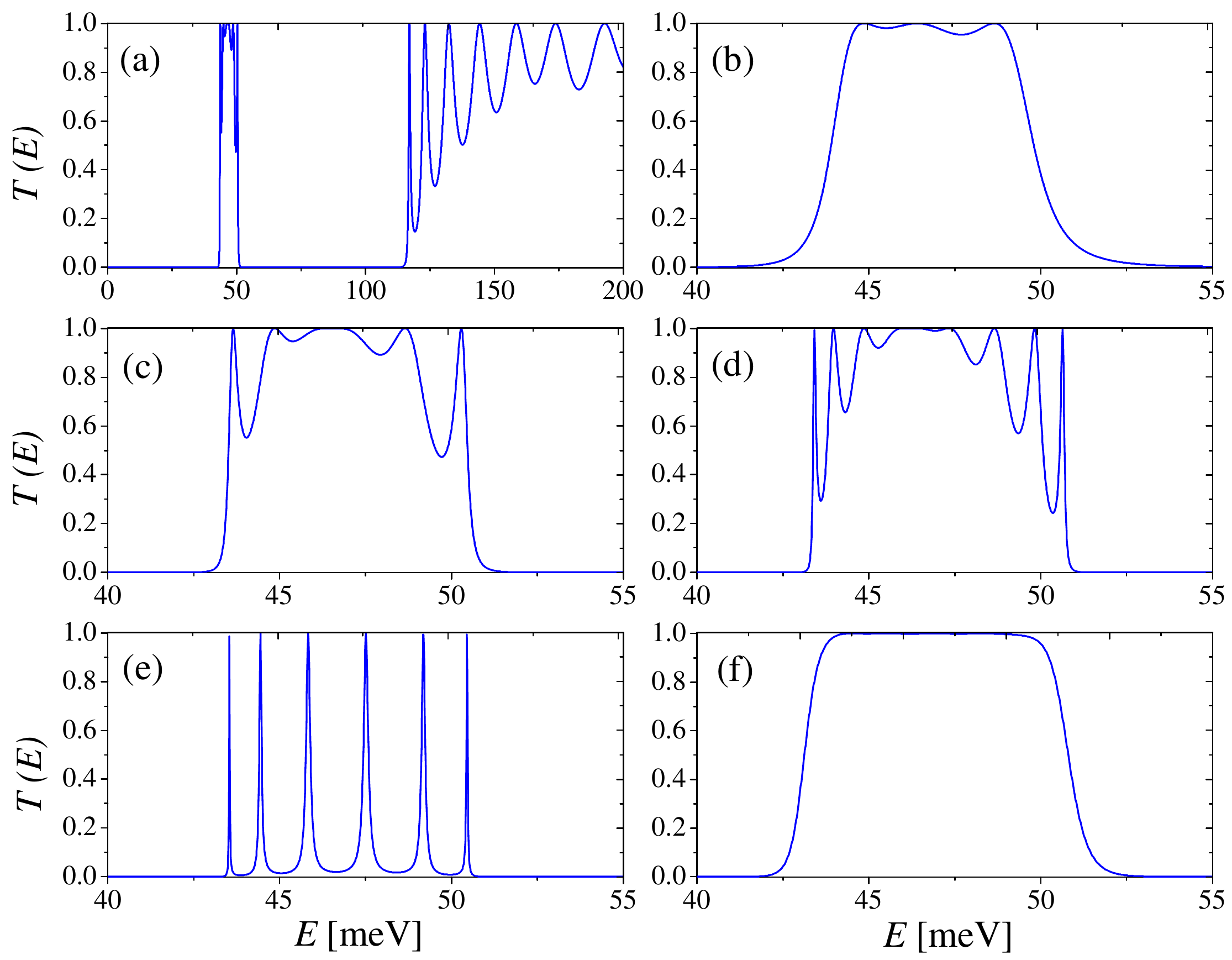}	
	\caption{\label{fig:transmission}(Color online) 
	(a) $T(E) = T(E, 0)$ plotted over a wide range of energies. 
	(b), (c) and (d) $T(E)$ close to the lowest transmission band, comparing 4, 7, and 10 barriers, respectively.
	(e) Same as (c) with 7 barriers, but without anti-reflection coating. 
	(f) $T(E)$ of 25 barriers where the width of the middle barrier is $15~\mathrm{nm}$ and the widths of other barriers 
	follows a Gaussian function (see main text). 
	}
\end{figure}

Figure~\ref{fig:transmission}(a) shows $T(E) = T(E,0)$ over a large range of energies. 
Below the barrier height transmission bands appear related to the minibands in an infinite superlattice
[there is only one such band with the parameters used in Figure~\ref{fig:transmission}(a)].
The widths of the transmission bands are determined by the tunnel amplitude 
through the barriers (determined primarily by $w$ and by $E$ relative to $\delta E_C$) and their spacings depend primarily on $d$.
Figures~\ref{fig:transmission}(b), (c), and (d) show a zoom of the lowest transmission band in Fig.~\ref{fig:transmission}(a) for 4, 7 and 10 
barriers, respectively and Fig.~\ref{fig:transmission}(e) compares with the case without anti-reflection coating which gives rise to a series of discrete
transmission peaks ill-suited for a thermoelectric device operating at a high power. With anti-reflection coating, adding more barriers 
makes the edges on either side of the 
energy bands steeper. Sharpening the lower edge of the transmission band prevents transport in the wrong direction as a response to a temperature difference and is 
crucial for high thermoelectric efficiency, but rather little is gained from adding more than 7 barriers.

An alternative path towards achieving a rectangular transmission function, the use of superlattices with a Gaussian distribution
of thicknesses or heights has been suggested~\cite{Gomez99}. Here we consider such a superlattice with 25 barriers, where the width of
barrier $i$ is $w \; \mathrm{exp}\{-[(i-13)/6]^2\}$. The corresponding transmission function is shown in Fig.~\ref{fig:transmission}(f)
and appears promising as it does not display any remainings of the isolated peaks. However, the onset is not as steep as for the case of the
superlattice with anti-reflection coating in Figs. 2(b) -- (d). Therefore, we found no
substantial improvement by using Gaussian superlattices, although we also tested other types
of variations, such as distance and barrier height. In addition, the large number of barriers needed in this case makes the transmission function 
very sensitive to random variations in the superlattice parameters.
 

\subsection{Power and efficiency of a power generator \label{sec:res_generator}}
\begin{figure}[]
  	\includegraphics[height=1.2\linewidth]{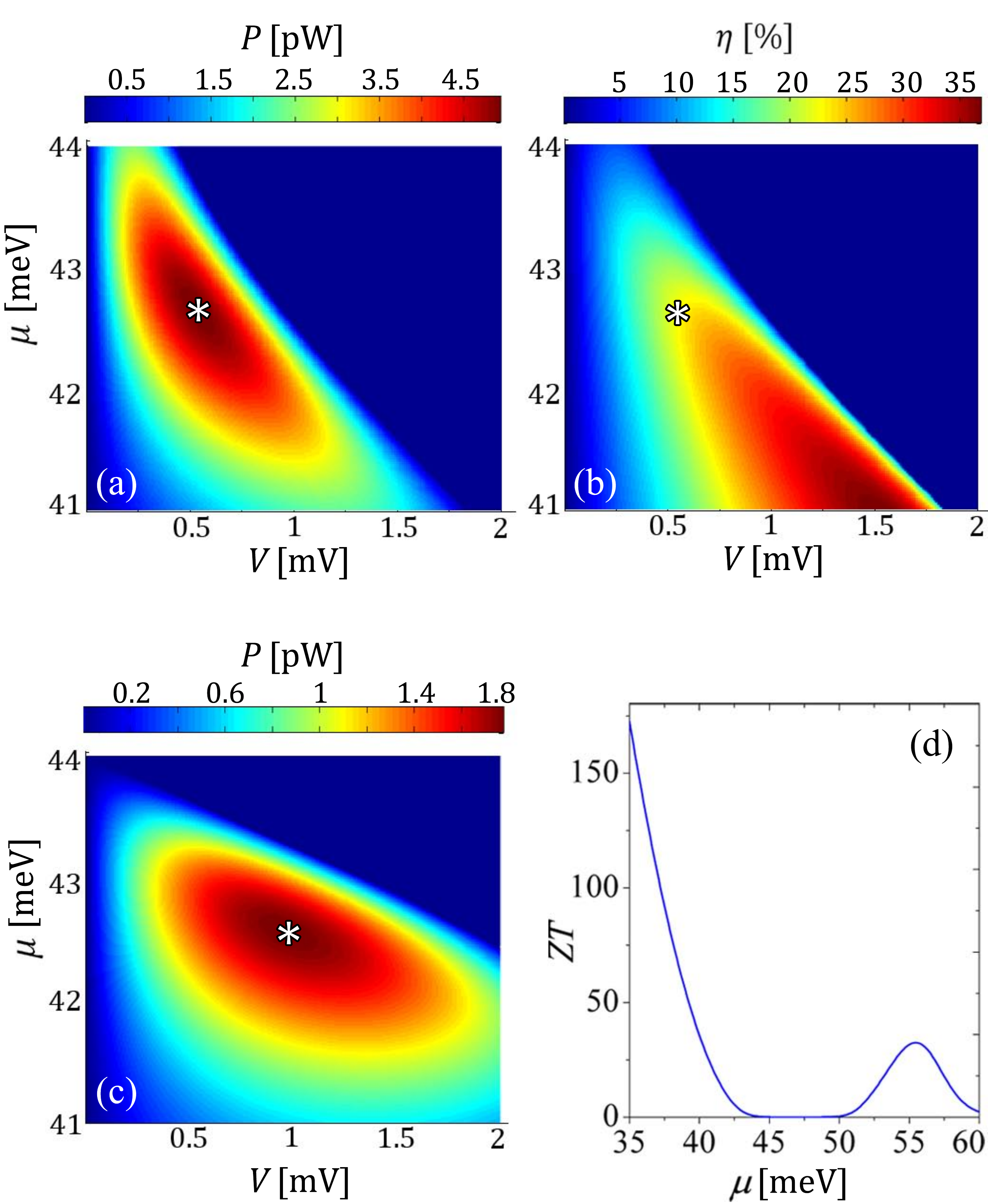}	
	\caption{\label{fig:efficiency_power}
	(Color online) Power $P$ (a) and efficiency $\eta$ (b)
	plotted on a color scale as a function of bias $V$ and average chemical
	potential $\mu$ of the leads for the 7-barrier superlattice with
	anti-reflection coating.  The white stars  mark the point
	of maximum output power. Note that negative values appearing for large
	$V$ are not displayed in our scale.  (c) $\eta(V,\mu)$	
	for a double barrier structure (single quantum dot) with barrier widths 
	adjusted to give the same efficiency at maximum
	power as the superlattice-structured NW in (b). (d) Figure of merit, ZT, as a function of $\mu$. }
%
	
\end{figure}

Figures~\ref{fig:efficiency_power}(a) and (b) show the output power and efficiency using the transmission function shown in Fig.~\ref{fig:transmission}(c),
plotted as functions of $V$ and the average lead chemical potential $\mu = (\mu_h + \mu_c)/2$. 
We focus on $\mu$ at or below the lowest transmission band, in which case thermoelectric transport is electron like; a similar result is found for 
$\mu$ at or above this transmission band, but then corresponding to hole-like thermoelectric transport.
$P$ and $\eta$ are here set to zero when the combination of $\mu$ and $V$ causes the current to flow from positive to negative biased leads and instead dissipate 
power. The maximum $\eta$ which can be achieved by appropriately tuning $V$ increases as $\mu$ falls below the lowest transmission band and 
can come arbitrarily close to $\eta_C = 60 \%$. However, from Fig.~\ref{fig:efficiency_power}(a) we see that the corresponding output power becomes very small 
in this region (Carnot efficiency can only be reached for reversible operation at vanishing output power~\cite{Humphrey02, Humphrey05}). 
A more suitable performance metric than the maximum efficiency is the efficiency at maximum power, $\eta_{maxP}$, which in Fig.~\ref{fig:efficiency_power}(a)
is $\eta_{maxP}=22.9\%$, while the maximum output power is $P_{max}=4.95~\mathrm{pW}$.

For comparison we also calculate the efficiency and output power of a quantum dot with a single orbital at energy $E_p$ described by the 
Lorentzian transmission function
\begin{equation}\label{eq:lorentzian}
	T_{QD}(E) = \frac{(\Gamma/2)^2}{(E-E_p)^2+(\Gamma/2)^2},
\end{equation}
independent of $V$. We adjust the width $\Gamma$ to obtain the same $\eta_{maxP}$ as for the superlattice-structured NW in 
Fig.~\ref{fig:efficiency_power}(b) and show the output power obtained with this value for $\Gamma$ in Fig.~\ref{fig:efficiency_power}(c). 
Although a Lorentzian transmission function is ideal to maximize $\eta$ and $\eta_{maxP}$ for small $\Gamma$, in the large $P$ regime of large $\Gamma$, 
it is seen to give almost a factor $3$ lower $P_{max}$ than the superlattice-structured NW. It is not at all possible 
to achieve a $P_{max}$ comparable with Fig.~\ref{fig:efficiency_power}(a) with a Lorentzian transmission function, no matter how large we make $\Gamma$.

The performance of thermoelectric devices is often characterized by the dimensionless thermoelectric figure of merit, $ZT = G S^2 T / \kappa$, where 
the conductance $G$, the Seebeck coefficient $S$, and the heat conductance $\kappa$ are given by linear-response versions of expressions like 
Eqs.~(\ref{eq:current}) and (\ref{eq:heatcurrent})~\cite{Sivan1986, Muller2008, Esfarjani2006} (leading order expansion in $V$ and $\Delta T = T_h - T_c$). 
When $ZT \rightarrow \infty$, $\eta \rightarrow \eta_C$, but this only holds in linear response and $ZT$ is not a particularly useful 
quantity for our study of the nonlinear regime. Nonetheless, for comparison we show in Fig.~\ref{fig:efficiency_power}(d) $ZT(\mu)$ plotted 
over a larger range of $\mu$ compared with Figs.~\ref{fig:efficiency_power}(a) and (b). For $\mu$ below the lowest transmission band $ZT$ grows 
to be very large, but as mentioned above this corresponds to the rather uninteresting regime of very small output power.
When $\mu$ is inside the transmission band, $ZT$ becomes small because the transmission is almost symmetric around $\mu$. For $\mu$ above the 
transmission band, thermoelectric transport becomes hole-like and $ZT$ grows again.


\subsection{Nonideal effects: Disorder and multiple subbands \label{sec:res_disorder}}
\begin{figure}[]
  	\includegraphics[height=0.99\linewidth]{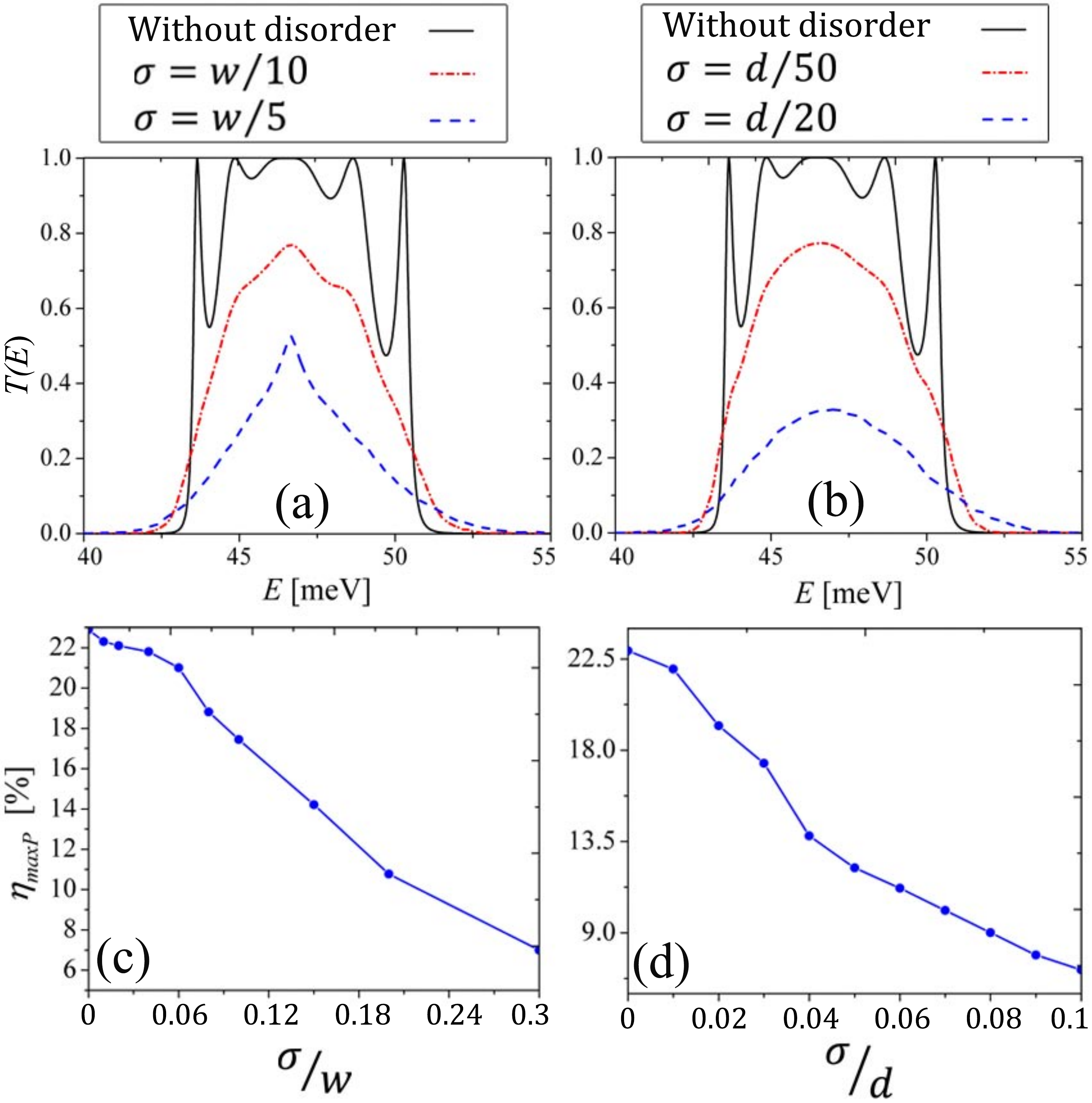}	
	\caption{\label{fig:disorder}(Color online) 
	(a) $T(E)$ with increasing disorder in $w$.
	(b) $T(E)$ with increasing disorder in $d$.
	(c) and (d) $\eta_{maxP}$ as a function of disorder strength in $w$ and $d$, respectively.
	}
\end{figure}

In this section we investigate deviations from the assumption of a strictly 1D NW with perfect barriers. First we introduce random variations 
into the widths and separations of the barriers. Random variations in barrier heights are likely small when determined by the 
crystal structure, but would have qualitatively the same effects as variations in widths and separations. We let the barrier widths and separations 
vary around their mean values according to a normal distribution with standard deviation $\sigma$, and calculate the average transmission 
function based on 5000 such randomly generated NWs [Figs.~\ref{fig:disorder}(a) and (b)]. Both types of variations reduce the 
height of the transmission function as well as the sharpness of the transmission band edges.
This average transmission function 
is then used to calculate $\eta_{maxP}$ [Figs.~\ref{fig:disorder}(c) and (d)], corresponding to the efficiency at maximum power which 
could be achieved in a thermoelectric device where those NWs were coupled in parallel. 

\begin{figure*}[]
  \includegraphics[height=5cm]{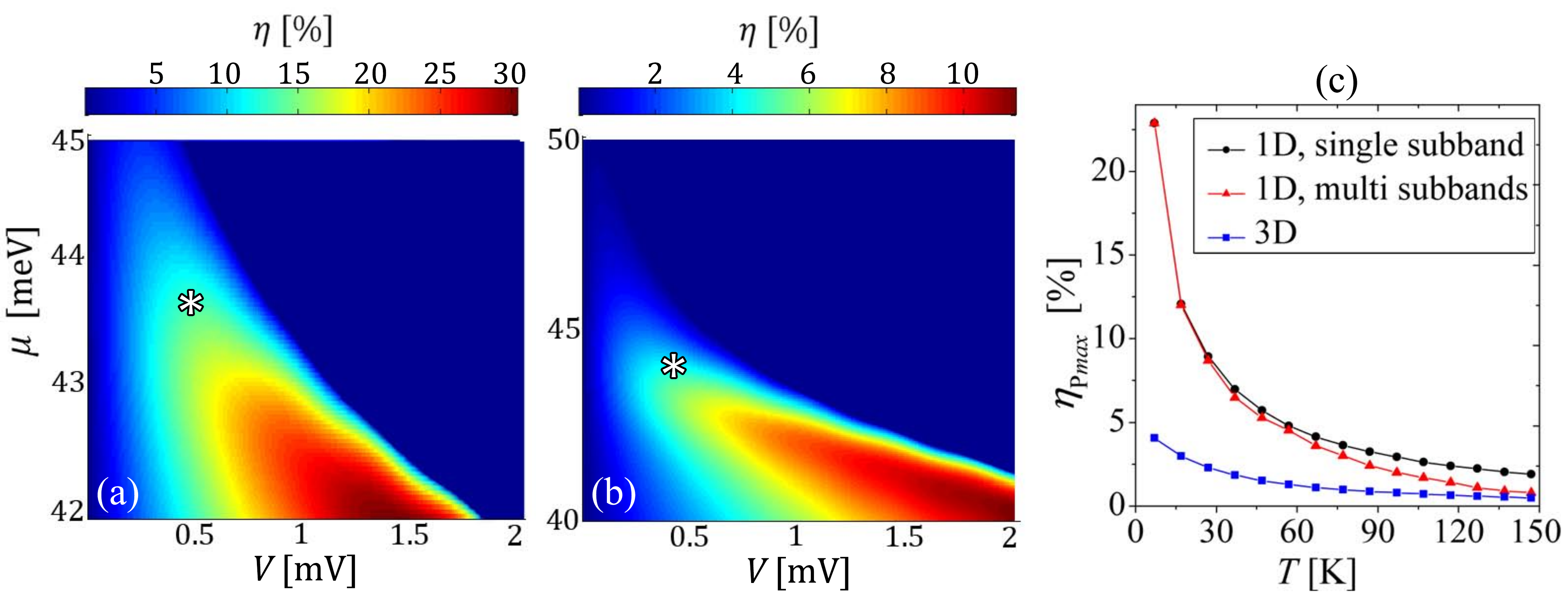}
\caption{\label{fig:multibands}(Color online)
	(a) $\eta(V,\mu)$ plotted on a color scale for a multi-subband NW with $R = 150~\mathrm{nm}$.
	(b) $\eta(V,\mu)$ plotted on a color scale for a bulk superlattice (equivalent to a NW with $R \rightarrow \infty$). 
	(c) $\eta_{maxP}$ as a function of average temperature for a single-band NW 
	($R \rightarrow 0$), a multi-subband NW ($R = 50~\mathrm{nm}$), and bulk ($R \rightarrow \infty$). 
	}
\end{figure*}

For the parameters used here, $\eta_{maxP}$ remains large for standard deviations up to around $10 \%$
of the average barrier width (or around $1.5$~nm), see Figs.~\ref{fig:disorder}(a) and (c). $\eta_{maxP}$ is more sensitive to variations in 
the barrier separation, see Figs.~\ref{fig:disorder}(b) and (d), decreasing substantially below standard deviations of around $2\%$ of the 
average barrier separation, meaning around $0.2$~nm, so here monolayer precision is desirable. 

We now take into account the finite width of the NW by including more than one 1D subband in the calculation. Here we make the simplest assumption that the 
NW has exactly the same cylindrical cross-section inside and outside the barriers and that the interfaces to the barriers are perfect, 
in which case there is no scattering between different 1D subbands and
the electric and heat currents are found from straightforward generalizations of Eqs.~(\ref{eq:current}) and (\ref{eq:heatcurrent})
\begin{eqnarray}\label{eq:current_multi}
	I &=& \frac{2 e}{h} \sum_n \int dE \; T_n(E, V) \nonumber \\ 
	  &\times& \left[ f\left(\frac{E-\mu_h}{k_B T_h}\right)-f\left(\frac{E-\mu_c}{k_B T_c}\right) \right],\\
\label{eq:heatcurrent_multi}
	Q_h &=& \frac{2}{h} \sum_n \int dE \; T_n(E, V) (E - \mu_h)  \nonumber \\ 
	    &\times& \left[ f\left(\frac{E-\mu_h}{k_B T_h}\right) - f\left(\frac{E-\mu_c}{k_B T_c}\right) \right],
\end{eqnarray}	 
where the transmission for the different subbands, $T_n(E,V)$, differ from each other only by the energy of the bottom of the different subbands, 
$T_n(E,V) = T_0(E - (E_n - E_0))$, where $E_n$ are the bottoms of the subbands, which are found by solving the 2D Schr{\"o}dinger 
equation in the circular NW cross-section using a single band effective mass approximation. 
For comparison we also calculate $P$ and $\eta$ for a superlattice-structured 3D material, equivalent to a NW with $R \rightarrow \infty$. 
In 3D, we need to integrate also over the transverse momenta when calculating the electric current density and heat current, resulting in~\cite{Daviesbook}

\begin{eqnarray}
\label{eq:current_3D}
	J^{3D} &=& \frac{e}{h} \int dE_{\perp} \; T(E_{\perp}, V) \nonumber \\ 
	  &\times& \left[ n_{2D}\left(\mu_h - E_{\perp}, T_c \right)-n_{2D}\left(\mu_c - E_{\perp}, T_h\right) \right],\\
\label{eq:heatcurrent_3D}
	Q_r^{3D} &=& \frac{m}{\pi^2 \hbar^3} \int dE_{\perp} \; T(E_{\perp}, V) \int dE_{\parallel} \; (E_{\perp}+E_{\parallel}-\mu_h) \nonumber \\ 
	    &\times& \left[ f\left( \frac{E_{\perp}+E_{\parallel}-\mu_h}{k_B T_h}\right) - f\left(\frac{E_{\perp}+E_{\parallel}-\mu_c}{k_B T_c}\right)\right], \nonumber \\ 
\end{eqnarray}	
where $E_{\perp}$ is the kinetic energy due to electron motion in the transport direction and $E_{\parallel}$ is the potential energy and kinetic energy in other 
directions, and $n_{2D}(\varepsilon, T) = \frac{m k_B T}{\pi \hbar^2}\mathrm{ln}(1+e^{\varepsilon/k_B T})$.

To see clear effects of the finite NW width we show in Fig.~\ref{fig:multibands}(a) $\eta$ for a rather thick NW with $R = 150$~nm, for 
which $\eta_{maxP}=12.01\%$, i.e., almost reduced by a factor 2 compared with the single-subband case in Fig.~\ref{fig:efficiency_power}(b).
Figure~\ref{fig:multibands}(b) shows $\eta$ for a 3D superlattice structure, which is even further reduced, with $\eta_{maxP}=4.07\%$.
In Fig.~\ref{fig:multibands}(c) we plot the calculated $\eta_{maxP}$ as a function of average lead temperature (keeping $T_h = T_c + 6~K$ constant),
comparing a superlattice-structured single-subband NW, a multi-subband NW with $R = 50$~nm, and a 3D structure. 
We see that a NW with $R = 50$~nm performs approximately as well as a perfect 1D system for $T \lesssim 30$~K, then looses in performance for higher $T$
as more subbands start to conduct, and finally approaches the performance of a 3D system for $T \gtrsim 150$~K.
At $T_h = 10$~K and $T_c = 4$~K, we find that a multi-subband NW retains more or less the high performance of a perfect 1D NW for $R \lesssim 70$~nm.

\section{Conclusions \label{sec:conclusions}}
In conclusion, we have investigated ballistic superlattice-structured NWs operated as thermoelectric generators, 
and have calculated the output power, efficiency, and efficiency at maximum power in the nonlinear regime with large temperature differences
between the hot and cold leads. Our results show that at low temperatures [$4$~K ($10$~K) for the cold (hot) lead was considered here], 
excellent performance can be achieved under conditions which should be within reach of present-day capabilities for epitaxial NW growth: 
rather few barriers ($\gtrsim 4$), relatively thick NWs ($R \lesssim 70$~nm), and with some 
tolerance for random variations in the barrier width and separation. 
The performance is much better than for a 3D superlattice-structured material,
and when operating at high output power, the efficiency is much larger than for a quantum dot (double-barrier structure).
For simplicity, we have focused on thermoelectric power generation, but the same criteria of efficient energy filtering of 
electrons will give rise to a high efficiency and output power also for a thermoelectric refrigerator.
\begin{acknowledgments}
We acknowledge stimulating discussions with M. Josefsson and H. Linke, and H. K. is grateful for the support of G. Rashedi and the 
Office of Graduate Studies at the University of Isfahan. We also acknowledge financial support from the Swedish Research Council (VR) and from the 
ministry of science research and technology of the Islamic republic of Iran.
\end{acknowledgments}
     

\end{document}